# Unveiling the amorphous ice layer during premelting using AFM integrating machine learning

Binze Tang,[1*], Chon-Hei Lo[2*], Tiancheng Liang[1*], Jiani Hong[1*†], Mian Qin[3], Yizhi Song[1], Duanyun Cao[4,5], Ying Jiang[1,6,7,8‡], Limei Xu[1,6,7§]

[1]*International Center for Quantum Materials, School of Physics, Peking University, Beijing, 100871, China*
[2] *Wenzhou Institute, University of Chinese Academy of Sciences, Wenzhou, Zhejiang 325001, China*
[3]*School of Physics, Peking University, Beijing, 100871, China*
[4]*Beijing Key Laboratory of Environmental Science and Engineering, School of Materials Science and Engineering, Beijing Institute of Technology, Beijing, 100081, China*
[5] *Chongqing Innovation Center, Beijing Institute of Technology, Chongqing, 401120, China*
[6]*Collaborative Innovation Center of Quantum Matter, Beijing, 100871, China*
[7]*Interdisciplinary Institute of Light-Element Quantum Materials and Research Center for Light-Element Advanced Materials, Peking University, Beijing 100871, China*
[8]*New Cornerstone Science Laboratory, Peking University, Beijing 100871, P. R. China*

Premelting plays a key role across physics, chemistry, materials and biology sciences but remains poorly understood at the atomic level due to surface characterization limitations. We report the discovery of a novel amorphous ice layer (AIL) preceding the quasi-liquid layer (QLL) during ice premelting, enabled by a machine learning framework integrating atomic force microscopy (AFM) with molecular dynamics simulations. This approach overcomes AFM's depth and signal limitations, allowing for three-dimensional surface structure reconstruction from AFM images. It further enables structural exploration of premelting interfaces across a wide temperature range that are experimentally inaccessible. We identify the AIL, present between 121-180K, displaying disordered two-dimensional hydrogen-bond network with solid-like dynamics. Our findings refine the ice premelting phase diagram and offering new insights into the surface growth dynamic, dissolution and interfacial chemical reactivity. Methodologically, this work establishes a novel framework for AFM-based 3D structural discovery, marking a significant leap in our ability to probe complex disordered interfaces with unprecedented precision and paving the way for future disciplinary research, including surface reconstruction, crystallization, ion solvation, and biomolecular recognition.

## I. INTRODUCTION.

Premelting—the formation of a thin liquid-like layer on crystal surfaces well below the melting point[1,2]—is a ubiquitous interfacial phenomenon observed across all classes of solids, with profound scientific and practical implications. It influences a wide range of properties and processes, including mechanical behavior, friction, chemical reactivity, cryopreservation, and atmospheric chemistry[3-8]. First proposed by Faraday on ice surface over 170 years ago, this phenomenon has been extensively explored through experimental and simulation techniques[9-13]. However, its underlying mechanism remains unsolved, primarily due to the challenge in probing the atomic structure and dynamics of disordered interfaces. Unlike bulk materials, which can be readily analyzed using crystallography[14,15], surface structures are inherently more complex and demand advanced, surface-sensitive techniques, such as low-energy electron diffraction[16], helium-atom scattering[9], X-ray absorption spectroscopy[17], and sum frequency generation spectroscopy[10,18-21]. While these methods yield valuable insights into the outermost layers, they suffer from limited spatial resolution and intrinsic averaging effects, which prevent the resolution of nanoscale heterogeneities.

Recent advancements in qPlus-based noncontact AFM (nc-AFM) with a CO-functionalized tip[22] has achieved submolecular resolution of surface structures[23-27], capturing ordered structures, transient intermediates, and even disordered configurations. Despite these capabilities, AFM faces fundamental challenges when applied to complex three-dimensional (3D) disordered systems. Conventional AFM analyses rely on trial-and-error workflows: candidate structures inferred from experimental images are relaxed using density functional theory (DFT), then compared to simulated AFM images via the probe particle method (PPM)[28,29]. While effective for simple, atomically flat surfaces[30-33], this approach becomes computationally prohibitive for structurally heterogeneous, fluctuating interfaces. Moreover, the intrinsic surface sensitivity of nc-AFM limits depth resolution, obstructing the reconstruction of full 3D structures and the understanding of interfacial dynamics[34,35].

\* These authors contributed to this work equally.

†Contact author: timeless@pku.edu.cn

‡ Contact author: yjiang@pku.edu.cn

§Contact author: limei.xu@pku.edu.cn

Machine learning (ML) offers new possibilities for AFM imaging interpretation, enabling advances in atomic identification[36-38], molecular classification[39-41], and electrostatic potential mapping[42]. However, current ML methods are predominantly designed for well-ordered or planar surfaces and struggle with disordered, non-periodic interfaces, where signal degradation, often compounded by experimental noise, leads to a pronounced increase in structural ambiguity. Although generative models have achieved success in areas such as protein structure prediction[43], organic molecules synthesis[44], and crystal[45,46] structure generation, robust 3D reconstruction of disordered and asymmetric interfaces from incomplete AFM data remains a formidable challenge[47].

Here we introduce a **general** ML-AFM framework that combines object detection for topmost layer structure identification with structure generation to infer subsurface configurations, enabling accurate atomic-scale reconstruction of disordered interfaces from AFM data. We apply this framework to ice premelting—the earliest and most extensively studied premelting system—and successfully reconstruct its disordered structure. The reconstructed 3D configurations provide physically grounded inputs for molecular dynamics simulations, enabling effective sampling of premelting dynamics in large temperature regime (>140 K) that are experimentally inaccessible due to desorption under vacuum conditions. Notably, our simulations reveal a previously unrecognized amorphous ice layer (AIL) that emerges prior to the formation of the quasi-liquid layer (QLL). This AIL, present within the 121–180 K range, features a disordered hydrogen-bond (HB) network with solid-like dynamics, challenging conventional views of interfacial premelting. This work establishes a generalizable strategy for resolving complex 3D disordered interfaces, enabling high-resolution imaging and structural discovery that yields new insights into interfacial phenomena across diverse process, including crystallization, ion solvation, molecular recognition, and heterogeneous catalysis.

## II. RESULTS

### A. Overview of the framework

We designed two networks: an object detection network and a structure generation network to resolve the 3D structure of interfacial ice from AFM data. The object detection network analyses the topmost layer experimental signals, while the structure generation network reconstructs the subsurface where no experimental signals are available as illustrated in Fig. 1. In the object detection task, a 3D U-Net-like[48] neural network (NN) takes AFM images as input and predicts the corresponding 3D structure represented by voxels containing position and species information (see methods and fig. S1). Once the top-layer structure is identified, a conditional variational auto-encoder (cVAE) is employed to generate the underlying ice structure for subsequent structural relaxation (see methods and fig. S3). Due to the lack of labeled experimental data, all training data is generated through simulations. MD simulations are used to explore the phase space of bulk ice interfaces, and the sampled structures are transformed into AFM images using the PPM. To replicate experimental noise, a CycleGAN is trained using unlabeled experimental AFM images (see methods and fig. S2). The augmented simulated AFM images are then used to train the object detection model, while the sampled structures are directly employed to train the structure generation model.

### B. Input data preparation and augmentation

To explore the ice interface phase space, we performed molecular dynamics (MD) simulations of hexagonal ice (Ih) with the basal <0001> plane exposed to the vapor phase (Fig. 1a). Simulations spanned 160 K to 260 K to capture the entire premelting regime, where the first layer forms a quasi-liquid layer (QLL) near 180 K and the second layer melts above 254 K based on TIP4P/Ice water model[49,50]. To enrich the dataset and mimic ice growth, simulations were performed using both pristine and excess water deposited on interfaces (see methods). Structures were sampled using a 2.5 × 2.5 × 0.3 nm$^3$ sliding detection window (Fig. 1I) to capture interfacial disorder in the *xy*-plane, with the *z*-depth determined by the density distribution along the *z*-axis (see methods). Over 60,000 sampled structures were randomly divided into training (60%), validation (20%), and testing (20%) sets. Each structure was then converted into a stack of 10 simulated AFM images at varying tip-sample distances as input for the object detection NN (see methods and fig. S4). To address the challenge of detecting weak signals from deeper oxygen and hydrogen atoms amidst experimental noise, a CycleGAN was trained on unlabeled experimental AFM images. This approach aligns unpaired simulated and experimental AFM images by mapping them to a shared latent space, with training concluding upon convergence of the Fréchet Inception Distance [51], enabling the NN to be trained with realistic noise (see methods and fig. S5 for examples of images with or without CycleGAN).

### C. Structure identification for the topmost bulk ice interface

After training the NN for object detection with 36,000 data for over 10 epochs where the loss converges. The NN's prediction aligns well with reference data, even in the z-direction where water molecules exhibit significant fluctuations (see fig. S6). The prediction accuracy for the interfacial ice structure on the basal plane reaches nearly 100% for oxygen atoms and 99% for hydrogen atoms. To further assess the NN's generalization capability, predictions were extended to interfacial water structures on the prism I plane, which has a larger inter-layer distance and exhibits different pre-melting behavior compared to basal plane. Tested on over 10,000 structures across temperatures from 160 K to 260 K, the NN achieved 97% accuracy for oxygen atoms and 86% accuracy for hydrogen atoms (see fig. S6). The NN inferred the positions of hydrogen atoms based on patterns learned from the training data, but due to their weaker signal compared to oxygen and the structural differences between the basal and prism I planes, the prediction accuracy for hydrogen atoms in the prism I dataset decreased. Despite this, the NN demonstrates robust generalization capabilities for object detection tasks.

We then utilized the NN, trained exclusively on simulation data to analyze two experimental AFM images, each with a size $\sim 4 \times 4$ nm$^2$. One image features an interfacial superstructure recently observed at 121 K on hexagonal ice (Ih)[52]. This ($\sqrt{19} \times \sqrt{19}$) periodic superstructure, consisting of mixed Ih- and cubic (Ic)-stacking nanodomains which was uncovered by MD simulations, provides an excellent benchmark for evaluating the NN's performance. The second image depicts a disordered interfacial structure on Ih at 135 K. The experimental images were divided into $2.5 \times 2.5$ nm$^2$ sections, and the NN predicted atom positions for each section. By combining these predictions, we reconstructed the final structures that correspond to the experimental AFM images (see Fig. 2 and the fig. S7 for origin predictions). Prediction errors are generally observed in regions with weaker AFM signals, particularly where O and H atoms are less prevalent. These discrepancies can be easily corrected through physical rules adjustments (see details in methods).

We then use the point charge PPM to simulate the AFM images based on the modified structure without the substrate. As shown in Fig. 2, the simulated AFM closely match the experimental data. The NN accurately predicts the positions of dangling OH bonds (with O-H bonds pointing obliquely upward toward the surface). Comparing the NN predictions with the superstructure benchmark, the NN achieves an accuracy of 94% for O for over 143 water molecules. It is worth noting that the NN's performance deteriorates when CycleGAN-generated data is not included (see fig. S8 and table S1). These results demonstrate the transferability and robustness of our object detection NN.

### D. Structure generation and relaxation

Although simulated AFM images closely resemble experimental ones, completing the underlying layer and verifying its stability through relaxation is crucial before making definitive assessments of NN predictions. In bulk ice at low temperatures, the internal structure forms a near-perfect hexagonal crystal. However, aligning the lower-layer structure with the upper-layer disordered structure through simple rotation, translation, and energy calculations is labor-intensive and prone to multiple local energy minima due to proton disorder. The delicate HB network can be destabilized by slight lattice misalignments, tilts, or dangling molecules in the interfacial. To address this, we first extended the voxel representation and sliding detection window to $2.5 \times 2.5 \times 0.9$ nm$^3$, enabling the object detection NN to handle both detection and generation tasks. However, tests using datasets from both the basal and prism I planes revealed a significant drop in prediction accuracy with increasing detection depth, correlating with weaker AFM signals[35] (see fig. S6).

To address this issue, we developed a cVAE to reconstruct the lower-layer structure of the bulk ice interface based on the upper-layer structure (see network details in the methods). The process involved encoding the lower-layer structure into a latent space, then decoding it according to the upper-layer structure. By sampling from the latent space distribution, the decoder generates new lower-layer structures. We used a large 3D voxel representation with dimension of $2.5 \times 2.5 \times 1.6$ nm$^3$ to retain the mid- to long-range disorder in the interfacial ice. Approximately 300 water molecules were generated during the process. To simplify the computation only oxygen atoms were retained in generation. Besides, we also included a published trajectory simulating using the coarse-grained mW model to obtain large grain boundaries in the topmost layer[52] (see methods). The model was trained for over 15 epochs, and the parameters with the lowest reconstruction loss were selected. Test with simulation data demonstrated that the NN accurately generated structures that closely matched the real structure, displaying hexagonal crystal morphologies after applying translation and rotation based on the upper-layer structure (see fig. S9).

To apply the model to experimental data, the interfacial structure was segmented into 2.5 nm sections, using each section as a conditional input to generate the corresponding lower-layer structure. Due to the non-periodic nature of the interfacial structure, we constructed the interfacial ice structure by matching the generated structures with a larger ice Ih template (Fig. 3a). After completing the crystal matching, we first fixed the upper-layer boundaries and performed preliminary relaxation of the hydrogen-bonds, followed by energy relaxation of the entire system (see simulations details in methods). The green lines in Fig. 3a and 3b show the relative atomic displacements before and after relaxation, while the root-mean-squared displacement (RMSD) distribution is depicted in Fig. 3c. The results indicate that most atomic displacements are under 1 Å, demonstrating the network's effectiveness in reconstructing the lower-layer structure. An RMSD value around 1.2 Å reflects the relative displacement of water molecules due to partial HB mismatches in the adjacent lower layer, without altering the HB topology of the interfacial water. During the energy relaxation of the disordered structure, an RMSD value around 2 Å suggests mismatches between lower and upper layers, disrupting the hydrogen-bond network of the interfacial water (Fig. 3d). This local planarization phenomenon, where additional water molecules appear in the center of octagonal rings, has been observed experimentally. Defects in the lower-layer structure may signal the onset of premelting[52]. The failure prediction of NN on small defects may stem from two main reasons. Firstly, the cVAE network lacks precision in predicting small defects in large system. More importantly, these structures were absent in the simulated system, highlighting a significant discrepancy between training and experimental data distributions. Despite the limited scope of the training data, the network performs well in predicting most disordered regions, showcasing strong generalization capability.

### E. Amorphous ice layer

At low temperatures (115 K–135 K), the dynamics of water molecules slow down significantly, and the timescales required for relaxation from random configurations to near-equilibrium states exceed the capabilities of conventional MD simulations[53]. In contrast, AFM experiments capture these slow dynamics over extended periods. Therefore, the 3D structures derived from experimental data by our ML framework serve as ideal initial configurations for MD simulations. This approach facilitates a comprehensive exploration of the structural and dynamic properties of water molecules in this challenging low-temperature regime, and enables the study of the premelting process at higher temperatures, which is otherwise inaccessible to direct AFM observation due to desorption in vacuum. Seven representative 3D structures, obtained from AFM experiments (115 K–135 K), containing 120–250 surface water molecules with lateral dimensions of 3.5–7 nm (see fig. S10), were used as starting configurations for MD simulations. Simulations were performed from 120 K to 240 K, with a 20 ns relaxation at each temperature (see methods). After relaxation, the surface structures exhibited a planar, topologically disordered HB network, rather than a defect-free hexagonal ice configuration. Notably, we observed that such 2D structure disorder is confined to the topmost layer of the surface, while artificially introduced defects in the subsurface layers are rapidly repaired during relaxation.

We used tetrahedrality and the proportions of six-membered rings to characterize the surface structure (see SI). Both parameters indicate low values and don't change significantly as temperature, suggesting a stable, highly disordered state between 120 K and 180 K (See Fig. 4a, b and S11). This contrasts significantly with previous simulation studies, which typically begin with a proton-disordered, pristine hexagonal surface and assume that topological defects gradually accumulate with increasing temperature (see Fig. 4b and S11 and S12). For instance, at 140 K, six-membered rings accounted for approximately 30% of the surface structure, a result that differs significantly from simulation based on a proton-disordered surface with minimal defects, or those simulating the deposition process, which exhibit 3D disorder with many ad-molecules (see Fig. 4a). This challenges the conventional assumption that surface disorder primarily arises only after the formation of the quasi-liquid layer above 180 K, driven by thermally activated diffusivity[1]. We identified a distinct **amorphous ice layer** phase preceding QLL formation, characterized by pronounced 2D topological disorder and solid-like dynamical properties. Further analysis of the distribution of surface dangling OH groups, compared to previous SFG experimental data[19](see SI), provides strong qualitative and quantitative agreement, offering compelling evidence for the existence of this newly identified surface phase (See Fig. 4e).

This finding necessitates a revision of the ice premelting phase diagram under high vacuum condition (Fig. 4b). We propose that at approximately 121 K, surface proton disorder and the boundary between Ic and Ih nanodomain facilitates the formation of vacancies. These vacancies reduce the binding strength between neighboring molecules, triggering a cascade of structural disordering that ultimately drives the transition from the ordered superstructure phase to the AIL phase[52]. While cascade disordering has previously been associated with QLL formation, a similar phenomenon—a cascade of structural disorder propagating across the proton-disordered surface of Ih ice—was proposed by Watkins[54] and observed in kinetic Monte Carlo simulations (0.11 µs) at 100 K[53]. As the temperature approaches ~180 K, thermal activation of surface diffusivity triggers in-plane particle diffusion, transforming the AIL phase into the QLL phase, resulting in a more fluctuating and disordered interface (Fig. 4d). Notably, the AIL phase exhibits significant heterogeneity in the degree of disorder, which can influence the onset temperature for molecular diffusivity (see methods and fig. S12 and S13).

Such surface heterogeneity suggests variations in adsorption energies[54], which could significantly impact crystal growth kinetics, dissolution dynamics, and catalytic activity[55]. For example, the presence of topological defects in AIL could significantly enhance the chemical reactivity of ice surfaces, with direct implications for heterogeneous chemistry in stratospheric clouds[56], where trace gases (e.g., $H_2O_2$, $SO_2$, HCl) undergo uptake and dissociative reactions[6,57-59]. This experimentally validated 3D structural information serves as a solid foundation for future explorations into the physical and chemical properties of interfacial ice.

## III. DISCUSSION

We introduce a robust machine learning framework capable of reconstructing the complex 3D atomic structure at disordered surfaces directly from experimental AFM data. Applied to the prototypical system of ice premelting, this approach reveals a previously unrecognized amorphous ice layer that forms prior to the quasi-liquid layer, thereby revising the phase diagram of ice and provides atomic-scale insights into the premelting process essential to cryospheric science, materials mechanics, atmospheric chemistry, and planetary phenomena.

Methodologically, the ability to resolve the amorphous ice layer underscores the power of our framework in revealing complex interfacial structures and advancing our understanding of disordered systems. The two-stage strategy, which decouples AFM analysis into object detection and 3D structure generation, mitigates the compounding errors arising from simulation artifacts, experimental noise, and AFM's limited depth resolution. Crucially, this framework bridges the longstanding gap between simulation and experiment: AFM images offer physically grounded initial configurations for molecular dynamics (MD) simulations, enabling the exploration of thermal and structural dynamics that are otherwise inaccessible.

Although developed in the context of ice, the proposed framework is broadly applicable to a wide range of nanostructure and disordered interfaces. By integrating with established generative approached—such as molecular design and crystal structure prediction[43,44] —it can be extended to diverse systems, including organic adsorbates, heterogeneous catalytic surfaces, and biomolecular assemblies. Furthermore, integration this strategy with complementary modalities, such as spectroscopy, tomography, or coherent X-ray imaging, could enable a more comprehensive understanding of interfacial structure and function. By resolving a long-standing enigma in ice premelting and enabling nanoscale 3D reconstruction of disordered interfaces, our work establishes a generalizable paradigm for AFM-guided analysis, with broad implications for interfacial science and the inverse design of functional materials.

## APPENDIX A: AFM EXPERIMENTS

All the experiments were performed with a combined noncontact AFM/STM system at 5 K using a home-made qPlus sensor equipped with a tungsten (W) tip (spring constant, $k_0 \approx 1,800$ N·m$^{-1}$; resonance frequency, $f_0 = 30.4$ kHz; quality factor, $Q \approx 100,000$). All AFM data were measured at 5 K under ultrahigh vacuum ($<3\times10^{-10}$ mbar). The AFM frequency shift ($\Delta f$) images were obtained with the CO-functionalized tips in the constant-height mode, respectively with 200 pm oscillation amplitude. The tip height in AFM imaging refers to the maximum tip height (set as 0 pm) during the height-dependent imaging process, at which the contrast of H-up water molecules can be clearly distinguished. Only the relative heights between images have a certain reference value. Image processing was performed by Nanotec WSxM.

## APPENDIX B: MD SIMULATIONS

To explore the phase space of the studied system, we simulated the bulk ice interface by constructing a hexagonal ice (Ih) structure with the basal face (<0001> plane) exposed to the vapor phase. The Tip4p/Ice force field was used in all simulations[50]. An 8-bilayer ice-Ih slab with dimensions of 10.61 nm × 9.19 nm was initially created by GenIce[60] package. To account for thermal expansion, we first ran simulations of these bulk ice configurations in a constant pressure canonical ensemble across a temperature range from 160 K to 260 K, at a pressure of 0 bar for 2 ns. The equilibrium configurations were then cleaved by introducing a vacuum layer of approximately 50 Å. Periodic boundary conditions were applied in all three directions of the simulation box. To acquire the proton-disordered ice surface, we use a heating and annealing process based on Ref [61]. Specifically, all oxygen atoms are fixed while hydrogen atoms are heated to 1200 K and then annealed to 4 K over 2 ns. All MD simulations were carried out using the Large-scale Atomic/Molecular

Massively Parallel Simulator (LAMMPS) package[62].

To enrich the dataset, we conducted simulations with two initial configurations: i) a pristine hexagonal ice interface, and ii) an interface with excess water molecules deposited to mimic the ice growth process. For both configurations, MD simulations were performed over a temperature range of 160 K to 260 K (in steps of 20 K) to sufficiently sample phase space. The bottom four layers were fixed for all simulations. For initial configuration i), the simulation timestep was 1 fs and the total simulation time at a given ($N$, $V$, $T$) is 10 ns where the first 2 ns were run for structure relaxation. For initial configuration ii), 2600 water molecules (equivalent to 1.5 bilayers) were deposited on the surface at a rate of 1 molecule every 2 ps. After deposition, simulations were conducted at a given ($N$, $V$, $T$) for 15 ns where the first 5 ns were run for structure relaxation. Additionally, to diversify the test data, simulations were also conducted on an Ih slab with the Prism I plane (<$10\bar{1}0$> plane) exposed to the vapor phase. During the structure generation neural network training, we also incorporated a previously published trajectory simulating deposition on a larger surface (352.2 Å × 366.1 Å) with extended relaxation time, using the mW model to obtain grain boundaries in the topmost layer[52].

## APPENDIX C: STRUCTURE SAMPLING

We extracted the configurations by sliding a probing window along simulation trajectories to prepare the data for model training. We set the detection window (Fig. 1, the first panel) in size of 2.5 × 2.5 × 0.3 nm$^3$, which is large enough to capture the disordered feature of interfacial water structures in the XY plane. Thus, all structures are restricted within the windows, and their corresponding simulated AFM images have the same size of 2.5 × 2.5 nm$^2$. The depth of the detection window in the normal direction is chosen based on the density distribution peak along the z-axis to capture sufficient information on hydrogen bonds and ring structures. Then, the window slides in the direction parallel to the substrate. In the XY plane, the window takes 2 steps with a stride of 0.5 nm. The configurations used for data acquisition were sampled from the MD simulations trajectories with a time interval of 1 ns. For the structure generation task, the detection window was extended to 2.5 × 2.5 × 1.6 nm$^3$. We note that the detection window can also rotate in the x-y plane to mimic the random orientation of the substrate.

## APPENDIX D: SIMULATIONS OF AFM IMAGES

The AFM images were simulated using a molecular mechanics model based on methods described in refs [28,29]. We performed AFM simulations to model the CO-tip based on the probe-particle model with the following parameters, effective lateral stiffness k = 0.50 N/m, atomic radius Rc = 1.661 Å, and Q = -0.05 e (e is the elementary charge). The parameters $r$ (van der Waals radius) and $\epsilon$ (potential well depth) of the Lennard–Jones pairwise potentials for the O and H atoms used in AFM simulations are: $r_H = 1.487$ Å, $\epsilon_H = 0.680$ meV, $r_O = 1.661$ Å, $\epsilon_O = 9.106$ meV. To reduce computation cost, the charge distribution is modeled as a point charge on each atom with $q_H = 0.4238$ e and $q_O = -0.8476$ e. These parameters can effectively reproduce most of the important features of experimental AFM images (see Fig. S4). We observed small changes in the simulation parameters for training data do not significantly change the predictions on the experimental data. The tip height in the AFM simulations is defined as the vertical distance between the metal tip apex and the topmost layer of the substrate. The oscillation amplitude of the tip in the simulated AFM images is 200 pm.

## APPENDIX E: 3D VOXEL REPRESENTATION

Inspired by the YOLO[63], we developed a refined 3D voxel representation to describe atomic structures. The space is divided into 32 × 32 × 4 cubic voxels, each with a space diagonal no longer than 1.5 Å, ensuring that no two identical atoms occupy the same voxel (Fig. 1 c). Each voxel can contain at most one oxygen and one hydrogen atom, represented by confidence scores $c_O$ and $c_H$ respectively, where $c_O, c_H \in [0,1]$. If an atom is present, its displacement relative to the voxel's lower left vertex is recorded as fractional coordinates d$x$, d$y$, and d$z$. This representation allows continuously prediction of 3D atomic positions while minimizing computational resources. The loss function for object detection NN is the weighted sum of binary cross-entropy for atom presence and mean square error for relative displacements. After loss calculation, non-maximum suppression (NMS) is applied to further refine the predictions. To assess model performance, we constructed a confusion matrix that considers both atom types and positions.

## APPENDIX F: OBJECT DETECTION NEURAL NETWORK

We refer to 'Code availability' for full technical details and below provides a high-level summary of the object detection model architecture. The architecture of the detection network is based on 3D U-Net[48], and is mostly modified from denoising diffusion probabilistic models (DDPM) [64] (see architecture illustration in Fig. S1). It consists of three up-sampling and down-sampling layers. Each layer contains a double convolution with a residual connection. The network has skip connections between the up-sampling and down-sampling layer with the same resolution. In the lower two layers, an attention module is followed by the res-block. After passing through the former network, the results will be interpolated as the shape of a 3D voxel representation, and one residual block and a multilayer perceptron (MLP) are followed.

The optimization objection is a weighted sum of binary cross entropy (BCE) of confidence and mean square errors (MSE) of the fractional coordinates:

$$L_D = L_{BCE} + L_{MSE} \qquad (1)$$

where

$$L_{BCE} = \frac{1}{2N} \sum_{O,H} \sum_{i=1}^{N} [-w_c(p \cdot c_i \cdot \log \hat{c}_i + (1 - c_i) \cdot \log(1 - \hat{c}_i))] \qquad (2)$$

and

$$L_{MSE} = \frac{1}{2N} \sum_{O,H} \sum_{i=1}^{N} \left[\frac{w_{xy}}{2}\left((dx_i - \widehat{dx}_i)^2 + (dy_i - \widehat{dy}_i)^2\right) + w_z(dz_i - \widehat{dz}_i)^2\right] \qquad (3)$$

where $w_c = 1.0, w_\rho = 0.5, w_z = 0.5$ are weighting factors. $c_i$, $dx_i$, $dy_i$ and $dz_i$ are the data labels of confidence, fractional coordinates in x, y and z directions, respectively. And $\hat{c}_i$, $\widehat{dx}_i$, $\widehat{dy}_i$ and $\widehat{dz}_i$ are the network's output. To balance the positive and negative samples, pos weight $p = 5$ is applied to the BCE term.

To eliminate redundant atoms from the neural network (NN) predictions, we applied a non-maximum suppression (NMS) algorithm based on atom-pair distances. Unlike traditional methods, the computation of Intersection over Union (IoU) is not used. Voxels with $c_i > 0.5$ are selected in descending order, and any surrounding voxels within a distance of $r_{NMS} < 2.0\text{Å}$ from the selected voxel are discarded. After removing invalid atoms, the confusion matrix is generated by pairing atoms within a cutoff distance of $r_M = 1.0\text{Å}$. This matrix is used to determine true positives (TP), false positives (FP), and false negatives (FN). The F1-score, quantifying the NN's performance, is then defined as:

$$F_1 = \frac{2TP}{2TP+FP+FN} \qquad (4)$$

A higher F1-score indicates better detection performance.

## APPENDIX G: CYCLEGAN

The CycleGAN includes two pair of identical generators and discriminators (see architecture illustration Fig. S2). We use a similar network architecture described in the object detection part but with fewer trainable parameters. The discriminator networks consist of three layers of conv-norm-activation blocks, followed by an MLP. CycleGAN enables style transfer between experimental and simulation data, and vice versa. The training process and the hyperparameters follow those used in the original paper[65]. The complete objective is to minimize the following loss function:

$$L = L_{GAN} + \lambda_1 \cdot L_{cyc} + \lambda_2 \cdot L_{idtenity} \qquad (5)$$

where $\lambda_1 = 10, \lambda_2 = 0.5$ are two hyperparameters. we use the Fréchet Inception Distance (FID) to estimate the NN performance. FID is computed using a pre-trained neural network to capture the image features[51]. The Inception v3 model with 2048 latent features is used for this purpose[66]. To adapt FID for 3D images, we treat the 3D image as a stack of 2D images.

## APPENDIX H: STRUCTURE GENERATION NEURAL NETWORK

The structure generation network is a conditional variational autoencoder[67], which includes two encoders and one decoder (see architecture illustration Fig. S3). Both encoders share the same architecture, which includes three down-sampling layers. Each layer features a residual block, and the latter two layers contain an extra attention block. The encoder $E_c$ encodes the interfacial layer into an 8-dimensional latent vector in Gaussian distribution $N(\boldsymbol{\mu_c}, \mathbf{I})$. Another encoder $E_{VAE}$ encodes the lower layer an 8-dimensional latent vector in Gaussian distribution $N(\boldsymbol{\mu}, \boldsymbol{\Sigma})$ where $\boldsymbol{\mu}$ is mean vector and $\boldsymbol{\Sigma}$ is log-variance vector. The decoder $\boldsymbol{D}$ generates the lower layer structure depending on resampling from $N(\boldsymbol{\mu}, \boldsymbol{\Sigma})$ during training process or $N(\boldsymbol{\mu_c}, \mathbf{I})$ during prediction process.

The training objective is to minimize the MSE between original data and reconstruction one along with the Kullback-Leibler (KL) divergence between the Gaussian distribution $N(\boldsymbol{\mu_c}, \mathbf{I})$ and $N(\boldsymbol{\mu}, \boldsymbol{\Sigma})$:

$$L = L_{BCE} + \gamma \cdot L_{MSE} + \beta \cdot L_{KL} \qquad (6)$$

where $L_{BCE}$ and $L_{MSE}$ are the same as in the detection stage. And KL divergence is

$$L_{KL} = -\frac{1}{2} \sum_{i}^{N} \left[1 + \Sigma_i + (\mu_i - \mu_{c,i})^2 - e^{\Sigma_i}\right] \qquad (7)$$

$\beta = 1.0$ is a hyperparameter introduced in beta-VAE[68]. $\gamma = 0.25$ is also a hyperparameter that allows the model to focus more on structural learning.

## APPENDIX I: STRUCTURE GENRATION AND RMSD RELAXATION

As shown in Fig. S6, the AFM simulated image based on the object detection network's raw prediction shows good alignment with experimental images. However, to further improve the accuracy of the structural model, regions where the two AFM images (experiment images and simulation images of raw prediction) deviated require manual adjustments, namely, adding or removing water molecules based on ice rules[69]. Such modifications were easy to implement and took minimal time. Following the adjustments, we fixed the dangling OH bonds and minimized the energy using the TIP4P/ice empirical potential. This workflow ensured the reliability of the initial structure used for subsequent structure generation.

After matching the crystal structure for the interfacial disordered ice based on the experimental AFM images, we first performed a relaxation of the hydrogen bonds. During this process, the oxygen atoms of the crystal substrate and the upper layer were fixed, while allowing the upper layer to drift as a whole. The hydrogen atoms are free. This relaxation was conducted in the NVT ensemble at 120 K for 2 ns. To be noticed, the same method was applied in the object detection process to adjust the hydrogen bonds where only the top layer exists. Following the hydrogen-bond adjustment, we fixed the boundaries of the upper layer and further minimized the energy for the entire system to validate the stability of the generated structure. The conjugate gradient (CG) algorithm was applied during energy minimization, which was set to stop when the largest force component on any atom was smaller than $1 \times 10^{-8}$ kcal/(mol·Å). The root-mean-squared displacement (RMSD) of the trajectories was tracked and only the free oxygen atoms in the upper layer were included in the calculation, as presented in Fig. 3c.

## APPENDIX J: MD SIMULATIONS TAKING GENERATED DATA AS INITIAL CONFIGURATIONS

The structure generated by the model does not consider proton ordering. To address this, proton ordering is adjusted through a heating and annealing process based on Ref [61]. Specifically, all oxygen atoms are fixed while hydrogen atoms are heated to 1200 K and then annealed to 4 K over 2 ns. Subsequently, the entire structure is relaxed at 20 K, with top-layer oxygen atoms constrained by virtual springs to preserve the topological structure. The spring constant is gradually reduced from 2 kcal/(mol·Å²) to 0.125 kcal/(mol·Å²) over five simulation steps spanning a total of 250 ps. The structure is then heated to 100 K over 2 ns without any constraints.

For further investigation of its topological ordering and premelting process at higher temperatures, the structure is incrementally heated to the desired temperature at a rate of 0.2 K/ps. The system is then relaxed for 20 ns, with data collection and analysis performed during the final 5 ns.

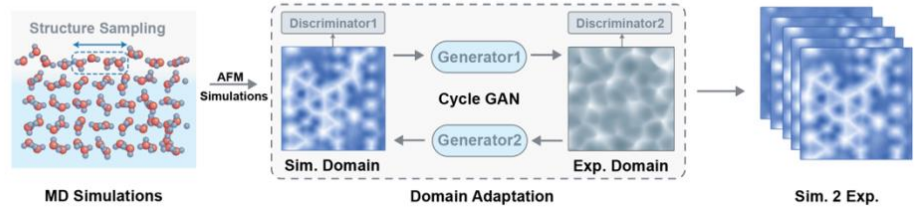

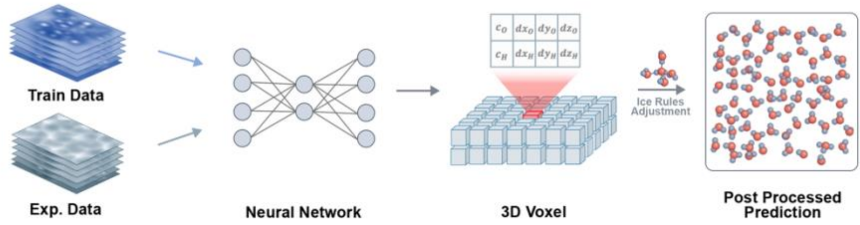

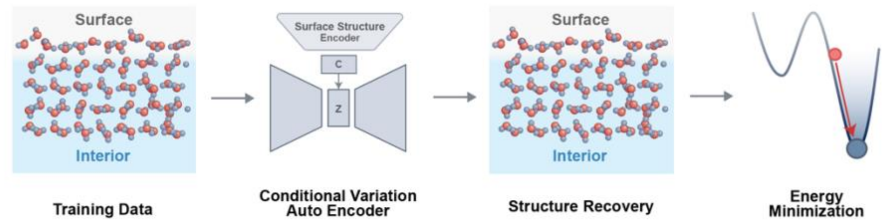

**Figure 1 | Schematic illustration of the overall framework of the training and prediction processes.** (Section I) Bulk ice surface structures are first sampled using MD simulations at temperatures ranging from 160 K to 260 K. These sampled structures are converted into AFM images via the PPM for training input data in Section II. Prior to training, a CycleGAN is trained on unlabeled experimental AFM images to introduce experimental-like noise into the simulated AFM images. (Section II) During training, a 3D U-Net-based neural network processes AFM images to predict the interfacial structure, represented by 3D voxels. For the prediction of experimental data, after adjustment based on ice rules, the predicted topmost layer structure can be simulated into AFM images without the substrate and relaxation for validation against experimental input. (Section III) To complete the underlying structure, a cVAE is trained using simulated data with the topmost layer structure as a conditional input. During prediction, the subsurface structure of the interfacial structure from Section II is reconstructed. The stability of the predicted structure is validated through energy minimization.

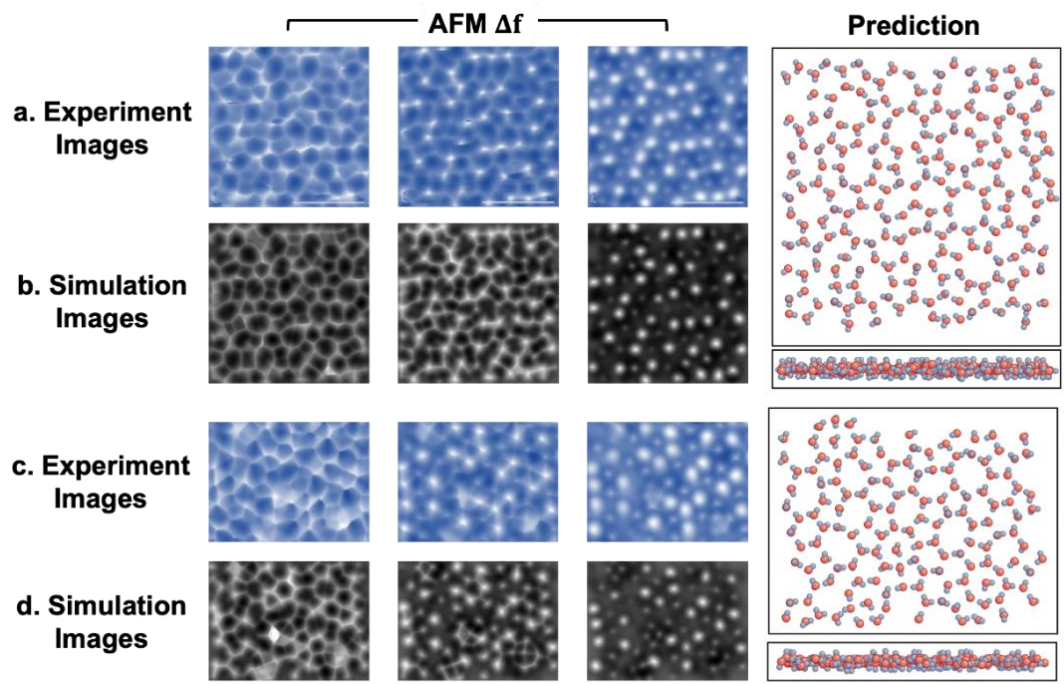

**Figure 2 | Examples of object detection network prediction from experimental data.** (a) and (c) Experimental AFM images of interfacial superstructure at 121 K and disordered structure at 135 K on the Ih phase basal plane. Columns 1–3 show AFM images at different tip heights -100 pm, -150 pm and -200 pm (from left to right). Column 4 shows the network prediction in top and side views, with red and white spheres representing oxygen and hydrogen atoms, respectively. (b) and (d) Simulated AFM images based on the network's predictions (no underlying structure included) corresponding to (a) and (c), showing increasing tip-sample distances.

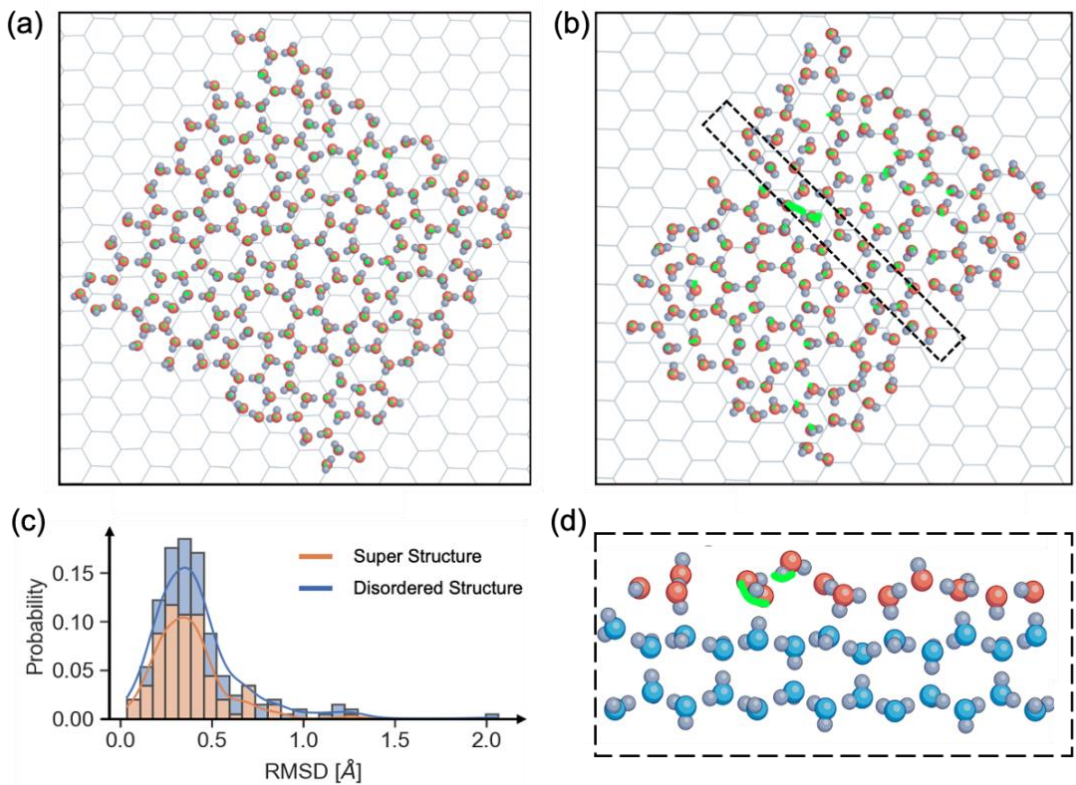

**Figure 3 | Prediction performance of structure generation network on experimental data.** (a) and (b) 3D structures of the interfacial superstructure and disordered structure from Fig. 2 (a) and (c), shown in top view. The underlying structures generated by the generative neural network are depicted in light gray lines. Green lines illustrate the relative atomic displacements of oxygen atoms in the top layers before and after relaxation. (c) RMSD distribution of oxygen atoms in the topmost layer during the energy relaxation process. The orange and blue lines represent the distributions for the superstructure in (a) and the disordered structure in (b), respectively. (d) Enlarged side view of the selected area within the dashed box in (b). Two of the largest displacements of oxygen atoms are indicated by green lines. Red, blue, and white spheres represent oxygen atoms in the top layer, lower layer, and hydrogen atoms, respectively.

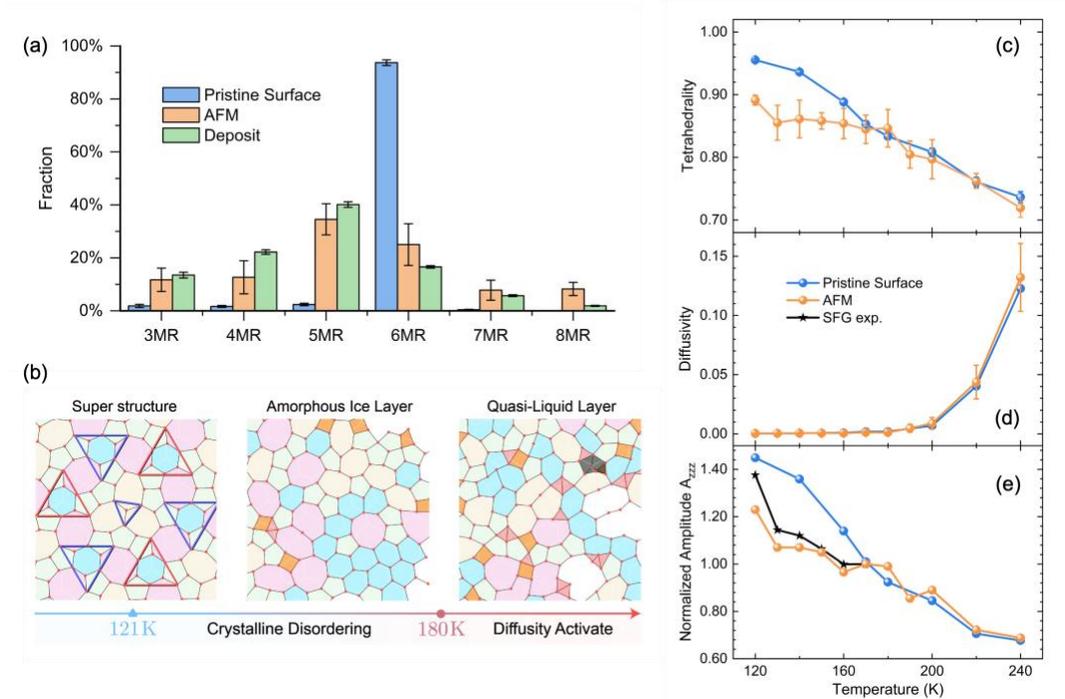

**Figure 4 | MD simulations using experimental data as initial configurations and Phase Diagram of the Ice Surface.** Distribution of hydrogen-bond ring sizes in the interfacial layer at 140 K for different initial configurations: disordered structure derived from atomic force microscopy, proton-disordered pristine Ih basal plane surface (Pristine Surface), and a deposition process (Deposit). Experimental results from AFM were averaged over 7 initial configurations, while results for the pristine surface and deposition process were averaged over three independent samples. (b) Phase diagram of the bulk ice surface under high vacuum conditions. A superstructure forms at 121 K; above this temperature, a structurally disordered amorphous ice layer (AIL) forms, transitioning to a quasi-liquid layer (QLL) around 180 K where molecular diffusivity is thermally activated. (c) Temperature dependence of the tetrahedral order parameter in the topmost layer. Orange and blue lines represent simulations initialized with configurations from AFM and the proton-disordered hexagonal surface, respectively. (d) Temperature dependence of the diffusion coefficient. (e) Temperature dependence of the sum-frequency generation (SFG) amplitude $A_{zzz} \propto N_s(0.32 \langle \cos\theta \rangle + 0.68 \langle \cos^3\theta \rangle)$ where $N_s$ and $\theta$ represent the number and polar orientation of dangling O-H, respectively, and $\langle \cdot \rangle$ denotes the ensemble average. The simulated $A_{zzz}$ values are normalized at 170 K based on the AFM values (orange dot). The experimentally observed spectral $A_{zzz}$ values are also normalized at 170 K[19].